\documentclass[aps,showpacs]{revtex4}
\usepackage{graphicx}% Include figure files
\usepackage{bm}% bold math
\begin{document}
\draft
\title{Eigenstates of the time-dependent density-matrix theory}
\author{Mitsuru Tohyama}
\address{Kyorin University School of Medicine, 
Mitaka, Tokyo 181-8611, Japan}
\author{Peter Schuck}
\address{
Institut de Physique Nucl$\acute{e}$aire, IN2P3-CNRS, Universit$\acute{e}$ Paris-Sud, F-91406 Orsay Cedex, France}
\date{\today}
\begin{abstract}
An extended time-dependent Hartree-Fock theory, known as the time-dependent density-matrix theory (TDDM), 
is solved as a time-independent eigenvalue problem
for low-lying $2^+$ states in $^{24}$O to understand the foundation of 
the rather successful
time-dependent approach.
It is found that the calculated strength distribution of the $2^+$ states
has physically reasonable behavior and that the strength function
is practically positive definite though the non-hermitian hamiltonian matrix obtained
from TDDM does not guarantee it.
A relation to an extended RPA theory with hermiticity is also investigated. 
It is found that the density-matrix formalism is a good
approximation to the hermitian extended RPA theory.
\end{abstract}

\maketitle

\section{Introduction}
The time-dependent density-matrix theory (TDDM) \cite{GT90} is a version of extended time-dependent
Hartree-Fock theories which dynamically include two-body correlations. Although TDDM has been formulated
to deal with large amplitude collective motions \cite{GTR90,TU02}, it has also been applied to small amplitude oscillations:
Dipole and quadrupole giant resonances in stable nuclei \cite{Fabio,T99} and
low-lying states in unstable oxygen isotopes \cite{T01,T02}. Although the obtained results are quite 
reasonable, the foundation of TDDM for small amplitude motions is not clear in its time-dependent form. In this paper 
an eigenvalue problem of the Small amplitude limit of TDDM (STDDM) is solved for low-lying
$2^+$ states in $^{24}$O to better understand the time-dependent approach.
The hamiltonian matrix of STDDM is not hermitian as will be shown below.
STDDM is compared with an extended RPA (ERPA) theory which has hermiticity \cite{TS} and its relation to the ERPA theory
is clarified.
This paper is organized as follows: The formulation of STDDM is presented in Sect. 2. A numerical solution
of STDDM for low-lying $2^+$ states in $^{24}$O is shown in Sect. 3. A comparison of STDDM with a more 
elaborate ERPA with hermiticity is made in Sect. 4 and Sect. 5 is devoted to a summary.

\section{Small amplitude limit of TDDM}
TDDM gives the time-evolution of
the one-body density-matrix $\rho(1,1')$ and the correlated part $C(12,1'2')$ 
of the two-body 
density-matrix, where numbers denote space, spin, and isospin coordinates.
The equations of motion for $\rho$ and $C$  have been obtained by truncating the
well-known Bogoliubov-Born-Green-Kirkwood-Yvon hierarchy of reduced density matrices \cite{WC}.
STDDM has been formulated according to the following steps \cite{TG1}:
1) Linearizing the equations of motion for $\rho$ and $C$ with respect
$\delta\rho$ and $\delta C$, where $\delta\rho$ and $\delta C$
denote deviations from the ground-state values $\rho_0$ and $C_0$ i.e.
$\delta\rho=\rho-\rho_0$ and
$\delta C=C-C_0$, respectively.
2) Expanding
$\delta\rho$ and $\delta C$ with single-particle states
$\psi_{\alpha}$ as
\begin{eqnarray}
\delta\rho(11',t)&=&\sum_{\alpha\alpha'}x_{\alpha\alpha'}(t)\psi_{\alpha}(1,t)
\psi_{\alpha'}^{*}(1',t), 
\\
\delta C(121'2',t)
&=&\sum_{\alpha_1\alpha_2\alpha_1'\alpha_2'}X_{\alpha_1\alpha_2\alpha_1'\alpha_2'}(t)
\psi_{\alpha_1}(1,t)\psi_{\alpha_2}(2,t)
\psi_{\alpha_1'}^{*}(1',t)\psi_{\alpha_2'}^{*}(2',t),
\end{eqnarray}
where $\psi_{\alpha}$ is chosen to be an eigenstate of the mean field hamiltonian $h_0(\rho_0)$:
\begin{eqnarray}
h_0(\rho_0)\psi_{\alpha}(1)=-\frac{\hbar^2\nabla^2}{2m}\psi_{\alpha}(1)+\int d2 v(1,2)
[\rho_0(2,2)\psi_{\alpha}(1)-\rho_0(1,2)\psi_{\alpha}(2)]=\epsilon_{\alpha}\psi_{\alpha}(1).
\label{hf}
\end{eqnarray}
Here the one-body density matrix $\rho_0$ is given as
\begin{eqnarray}
\rho_0(11')&=&\sum_{\alpha\alpha'}n^0_{\alpha\alpha'}\psi_{\alpha}(1)
\psi_{\alpha'}^{*}(1').
\end{eqnarray}
After Fourier transforming the linearized equations,
the equations for $x_{\alpha\alpha'}(\omega)$ and $X_{\alpha_1\alpha_2\alpha_1'\alpha_2'}(\omega)$ 
can be written in matrix form,
\begin{eqnarray}
\left(
\begin{array}{cc}
a & c \\
b & d
\end{array}
\right)
\left(
\begin{array}{c}
x \\
X
\end{array}
\right)
=\omega
\left(
\begin{array}{c}
x \\
X
\end{array}
\right),
\label{stddm1}
\end{eqnarray}
where $\omega$ is an eigenvalue.
The matrices $a, b, c,$ and $d$ are explicitly given in the Appendix.
Eq.(\ref{stddm1}) can also be obtained from the following equations:
\begin{eqnarray}
\langle\Phi_0|[a^+_{\alpha'}a_{\alpha},H]|\Phi\rangle
&=&\omega\langle\Phi_0|a^+_{\alpha'}a_{\alpha}|\Phi\rangle ,
\label{var1}
\\
\langle\Phi_0|[a^+_{\alpha_1'}a^+_{\alpha_2'}a_{\alpha_2}a_{\alpha_1},H]|\Phi\rangle
&=&\omega\langle\Phi_0|a^+_{\alpha_1'}a^+_{\alpha_2'}a_{\alpha_2}a_{\alpha_1}|\Phi\rangle,
\label{var2}
\end{eqnarray}
where 
$[~]$ stands for the commutation relation, $H$ is the total hamiltonian
consisting of the kinetic energy term and a two-body interaction, $|\Phi_0\rangle$ 
is the ground-state wavefunction and
$|\Phi\rangle$ the wavefunction for the excited state with excitation energy $\omega$.
Linearizing Eqs.(\ref{var1}) and (\ref{var2})
with respect to $x_{\alpha\alpha'}=\langle\Phi_0|a^+_{\alpha'}a_{\alpha}|\Phi\rangle$
and $X_{\alpha\beta\alpha'\beta'}=
\langle\Phi_0|a^+_{\alpha_1'}a^+_{\alpha_2'}a_{\alpha_2}a_{\alpha_1}|\Phi\rangle$, and
using the occupation matrix
$n^0_{\alpha\alpha'}=\langle\Phi_0|a^+_{\alpha'}a_\alpha|\Phi_0\rangle$ and
the two-body correlation matrix
$C^0_{\alpha_1\alpha_2\alpha_1'\alpha_2'}=
\langle\Phi_0|a^+_{\alpha_1'}a^+_{\alpha_2'}a_{\alpha_2}a_{\alpha_1}|\Phi_0\rangle
-{\cal A}(n^0_{\alpha_1\alpha_1'}n^0_{\alpha_2\alpha_2'})$,
where ${\cal A}$ is an antisymmetrization operator,
we arrive at Eq.(\ref{stddm1}). Eq.(\ref{stddm1}) is reduced to the second RPA (SRPA) \cite{Saw,Wam} under the two
assumptions: The one is the Hartree-Fock (HF) approximation for the ground state, in which
$n^0_{\alpha\alpha'}=\delta_{\alpha\alpha'}f_\alpha$ with $f_\alpha=1~(0)$ 
for occupied (unoccupied) states at zero temperature
and $C^0_{\alpha_1\alpha_2\alpha_1'\alpha_2'}$=0. The other is to restrict $X_{\alpha_1\alpha_2\alpha_1'\alpha_2'}$
only to two particle - two hole components and their complex conjugates.

The hamiltonian matrix of Eq.(\ref{stddm1}) is not hermitian, i.e. $b\neq c^+$, as is easily understood from its
explicit form.
For a non-hermitian hamiltonian matrix, the left-hand-side eigenvectors of the
hamiltonian matrix constitute a basis which is orthogonal to $(x_{\alpha\alpha'}, X_{\alpha_1\alpha_2\alpha_1'\alpha_2'})$,
and
the ortho-normal condition is written as
\begin{eqnarray}
\langle\tilde{\mu}|\mu'\rangle&=&
\sum_{\alpha\alpha'}\tilde{x}^*_{\alpha\alpha'}(\mu)x_{\alpha\alpha'}(\mu')
+\sum_{\alpha_1\alpha_2\alpha_1'\alpha_2'}\tilde{X}^*_{\alpha_1\alpha_2\alpha_1'\alpha_2'}(\mu)
X_{\alpha_1\alpha_2\alpha_1'\alpha_2'}(\mu')
=\delta_{\mu\mu'},
\end{eqnarray}
where $|\mu\rangle$ represents an eigenvector $(x_{\alpha\alpha'}, X_{\alpha_1\alpha_2\alpha_1'\alpha_2'})$
with the eigenvalue $\omega_{\mu}$, and $|\tilde{\mu}\rangle$ the left-hand-side eigenvector of
the hamiltonian matrix with the same eigenvalue:
\begin{eqnarray}
(\tilde{x}^*~~\tilde{X}^*) 
\left(
\begin{array}{cc}
a & c \\
b & d
\end{array}
\right)
=\omega_\mu
(
\tilde{x}^*~~\tilde{X}^*
) .
\label{stddm2}
\end{eqnarray}
The completeness relation is written as
\begin{eqnarray}
\sum_{\mu}
\left(
\begin{array}{c}
x_{\alpha\alpha'}(\mu)\\
X_{\alpha_1\alpha_2\alpha_1'\alpha_2'}(\mu)
\end{array}
\right)
(\tilde{x}^*_{\beta\beta'}(\mu)~\tilde{X}^*_{\beta_1\beta_2\beta_1'\beta_2'}(\mu))
=I,
\label{comp}
\end{eqnarray}
where $I$ is the unit matrix. In general
the strength function is defined as
\begin{eqnarray}
S(E)&=&\sum_{E_\mu>0}|\langle\Psi_\mu|\hat{Q}|\Psi_0\rangle|^2\delta(E-E_\mu),
\end{eqnarray}
where $|\Psi_0\rangle$ is the ground-state, $|\Psi_\mu\rangle$ is an excited state 
with an excitation energy $E_\mu$, and $\hat{Q}$ an excitation operator.
The strength function in STDDM is calculated in a way similar to that used in TDDM.
In TDDM, an initial excited wavefunction $|\Phi(t=0)\rangle$ is given by boosting the ground-state
wavefunction $|\Phi_0\rangle$ as
\begin{eqnarray}
|\Phi(t=0)\rangle=e^{ik\hat{Q}}|\Phi_0\rangle.
\end{eqnarray}
Then the initial values of the time-dependent amplitudes $ x_{\alpha\alpha'}$ and $X_{\alpha_1\alpha_2\alpha_1'\alpha_2'}$ 
are given 
at first order of $k$ as:
\begin{eqnarray}
x_{\alpha\alpha'}(0)&=&\langle\Phi(t=0)|a^+_{\alpha'}a_{\alpha}|\Phi(t=0)\rangle
\approx ik\langle\Phi_0|[a^+_{\alpha'}a_{\alpha},\hat{Q}]|\Phi_0\rangle
\\
X_{\alpha_1\alpha_2\alpha_1'\alpha_2'}(0)&\approx& 
ik\langle\Phi_0|[a^+_{\alpha_1'}a^+_{\alpha_2'}a_{\alpha_2}a_{\alpha_1},\hat{Q}]|\Phi_0\rangle.
\end{eqnarray}
In TDDM 
the time-dependence of $x_{\alpha\alpha'}$ and $X_{\alpha_1\alpha_2\alpha_1'\alpha_2'}$ is determined
by solving numerically the TDDM equations. It can also be
expressed using the eigenstates of Eq.({\ref{stddm1}}) as
\begin{eqnarray}
\left(
\begin{array}{c}
x(t) \\
X(t)
\end{array}
\right)=\exp
\left[-i
\left(
\begin{array}{cc}
a & c \\
b & d
\end{array}
\right)t/\hbar
\right]
\left(
\begin{array}{c}
x(0) \\
X(0)
\end{array}
\right)
=\sum_\mu \exp\left[-i\frac{\omega_\mu t}{\hbar}\right]
\left(
\begin{array}{c}
x_\mu \\
X_\mu
\end{array}
\right)
(\tilde{x}^*_\mu~\tilde{X}^*_\mu)
\left(
\begin{array}{c}
x(0) \\
X(0)
\end{array}
\right),
\label{stddm3}
\end{eqnarray}
where the completeness relation Eq.(\ref{comp}) is inserted.
The time-dependent moment $Q(t)=\langle\Phi(t)|\hat{Q}|\Phi(t)\rangle$
for a one-body operator becomes
\begin{eqnarray}
Q(t)=\sum_{Re (\omega_\mu>0)}\left\{\exp\left(-i\frac{\omega_\mu t}{\hbar}\right)
\left(\sum_{\alpha\alpha'}\langle\alpha|Q|\alpha'\rangle x_{\alpha'\alpha}(\mu)\right)
(\tilde{x}^*_\mu x(0)+\tilde{X}^*_\mu X(0))+{\rm complex~conjugate}\right\},
\label{qt}
\end{eqnarray}
where we use the fact that both states with $\omega_\mu$ and $-\omega_\mu^*$ are eigenstates of
Eq.(\ref{stddm1}).
The strength function in TDDM is obtained from the Fourier transformation of $Q(t)$ \cite{GT90} as
\begin{eqnarray}
S(E)=\frac{1}{\pi k\hbar}\int_{0}^{\infty}Q(t)e^{-\Gamma t/2\hbar}\sin\frac{Et}{\hbar}
dt,
\label{se0}
\end{eqnarray}
where an artificial damping factor $\Gamma$ is introduced to obtain a smooth distribution for $S(E)$.
Equations (\ref{qt}) and (\ref{se0}) suggest that the strength function in STDDM has the form
\begin{eqnarray}
S(E)&=-&\frac{1}{\pi}Im\{
\sum_{ Re(\omega_\mu)>0}[\left(\sum_{\alpha\alpha'}\langle\alpha|Q|\alpha'\rangle x_{\alpha'\alpha}(\mu)\right)
\left(\sum_{\beta\beta'}\langle\beta|Q|\beta'\rangle\tilde{x}^t_{\beta'\beta}(\mu)\right)^*
\frac{1}{E-\omega_\mu+i\Gamma/2}
\nonumber \\
&-&
\left(\sum_{\alpha\alpha'}\langle\alpha|Q|\alpha'\rangle x_{\alpha'\alpha}(\mu)\right)^*
\left(\sum_{\beta\beta'}\langle\beta|Q|\beta'\rangle\tilde{x}^t_{\beta'\beta}(\mu)\right)
\frac{1}{E+\omega_\mu^*+i\Gamma/2}
]\},
\label{se}
\end{eqnarray}
where
$\tilde{x}^t_{\alpha\alpha'}(\mu)$ is defined as
\begin{eqnarray}
\tilde{x}^t_{\alpha\alpha'}(\mu)
=\sum_{\lambda\lambda'}\langle\Phi_0|[a^+_{\alpha'}a_{\alpha},a^+_{\lambda}a_{\lambda'}]|\Phi_0\rangle
\tilde{x}_{\lambda\lambda'}(\mu)
+\sum_{\lambda_1\lambda_2\lambda'_1\lambda'_2}
\langle\Phi_0|[a^+_{\alpha'}a_{\alpha},a^+_{\lambda_1}a^+_{\lambda_2}a_{\lambda_2'}a_{\lambda_1'}]|\Phi_0\rangle
\tilde{X}_{\lambda_1\lambda_2\lambda'_1\lambda'_2}(\mu).
\end{eqnarray}
The strength function $S(E)$ in STDDM is not guaranteed to be positive definite, as is easily understood
from its expression Eq.(\ref{se}).
 
\section{Numerical solution}

In this section we present a numerical solution of Eq.(\ref{stddm1}) calculated for low-lying $2^+$ states
in $^{24}$O, where a time-dependent approach has previously been applied \cite{T02}.
The $E2$ strength function is calculated 
according to the following 
steps:

1) A static HF calculation is performed to obtain
the initial ground state. The neutron 2$s_{1/2}$ state is assumed to be
the last fully occupied neutron orbit of $^{24}$O.
The Skyrme III 
is used as the effective interaction. 
It has been used as one of standard parameterizations of the Skyrme force in nuclear structure calculations 
even for very neutron rich nuclei \cite{Otsu,Khan2}.
The single-particle wavefunctions are confined to a sphere
with radius 12 fm. The mesh size used is 0.1 fm.

2) To obtain the correlated ground state $|\Phi_0\rangle$, we evolve the HF ground state 
using the TDDM equations for $\psi_\alpha(1,t)$, $n_{\alpha\alpha'}(t)$ and $C_{\alpha_1\alpha_2\alpha_1'\alpha_2'}(t)$
\cite{T01,T02} 
and the following time-dependent residual
interaction 
\begin{eqnarray}
v(t)=(1-e^{-t/\tau})v(\bm{r}-\bm{r}').
\end{eqnarray}
The time constant $\tau$ should be sufficiently large 
to obtain a nearly stationary solution of the TDDM equations \cite{T95}.
We choose $\tau$ to be 300 fm/$c$. 
In the calculation of
$n_{\alpha\alpha'}(t)$ and $C_{\alpha_1\alpha_2\alpha_1'\alpha_2'}(t)$,
the single-particle states are limited to the neutron orbits $1d_{5/2}, 2s_{1/2}$ and $1d_{3/2}$.
In a consistent calculation the residual interaction should be the same as 
that
used to generate the mean field. However, a Skyrme-type force contains momentum dependent terms, 
which make the computation time of two-body matrix elements quite large. Therefore,
we need to use a simple force of the $\delta$ function
form $v\propto\delta^3(\bm{r} -\bm{r}')$. 
We use the following pairing-type residual interaction of the density-dependent $\delta$ 
function form \cite{Chas}
\begin{eqnarray}
v(\bm{r}-\bm{r}')=v_{0}(1-\rho (\bm{r})/\rho_0)
\delta^3(\bm{r}-\bm{r}'),
\end{eqnarray}
where $\rho (\bm{r})$ is the 
nuclear density. The parameters $\rho_0$ and $v_{0}$ are set to be 0.16fm$^{-3}$ and $-1000$ MeV fm$^3$,
respectively. Similar values of $\rho_0$ and $v_{0}$ have been used in the Hartree-Fock-Bogoliubov
calculations
\cite{Terasaki,Duguet,Yamagami} in a truncated single-particle space.
The time step used to solve the TDDM equations is
0.75 fm/$c$.

3) At $t=5\tau$ we stop the TDDM calculation and solve Eq.(\ref{stddm1}). Since we are interested in
low-lying states which originate in inner-shell transitions, we use only the 
the neutron $1d_{5/2}, 2s_{1/2}$ and $1d_{3/2}$ states. 
The dimension of the hamiltonian matrix is about $650\times650$ when these single-particle states are used.
The strength function for $2^+$ states
is calculated using Eq.(\ref{se}) with $Q=r^2Y_{20}(\theta,\phi)$. 
We use $\Gamma=0.5$MeV.

\begin{figure*}
%  \begin{center}
%\resizebox{0.75\textwidth}{!}
    \includegraphics[height=7cm]{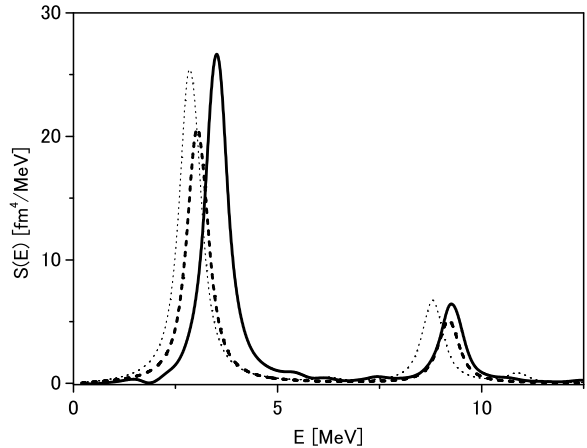}
%  \end{center}
  \caption{Strength distributions of the neutron quadrupole modes
in $^{24}$O
calculated in STDDM (solid line), RPA (dotted line) and SRPA (thin dotte line).}
\end{figure*}

The eigenvalues of some $2^+$ states become imaginary because the hamiltonian matrix of Eq.(\ref{stddm1}) is 
not hermitian. However, their imaginary components are quite small (less than 0.08MeV). 
Some $2^+$ states have also negative quadrupole strengths because positivity of $S(E)$ is not guaranteed.
However, the negative contributions are so small that $S(E)$ becomes almost positive in the entire energy region 
when it is smoothed with $\Gamma=0.5$MeV.
The obtained result in STDDM (solid line) is shown in Fig.1, 
where the strength functions in RPA (thick dotted line) and SRPA (thin dotted line) are also
presented for comparison. The first $2^+$ state calculated in STDDM
is energetically shifted upward and becomes significantly more collective 
as compared with that in RPA. The increase in the excitation energy is due to the lowering of the ground state, and
the enhancement of the collectivity of the first $2^+$ state is due to
the mixing of two-body configurations induced by ground-state correlations.
These properties of the first $2^+$ state under the influence of ground-state correlations
are similar to those obtained from quasi-particle RPA calculations \cite{Matsuo,Khan3} and also from
the TDDM calculation \cite{T02}. Large single-particle space including
continuum states has been used in these realistic calculations \cite{Matsuo,Khan3,T02} 
so as to deal with giant resonances as well as low-lying states.
Therefore, only a semi-quantitative comparison between our previous TDDM result \cite{T02} 
and the present STDDM calculation,
which can only be performed in very truncated
single-particle space, should be made for lowest-lying states.
In SRPA where ground-state correlations are neglected, the 2$^+$ states
become also collective due to the coupling to two particle - two hole configurations,
as shown in Fig.1. However, their excitation energies
are simply shifted downward. This might cause serious problems when unperturbed particle - hole energies
are small and (or) when the coupling to two-body configurations is strong.
The numerical solution of Eq.(\ref{stddm1}) shown in this section is physically reasonable and also
gives additional justification to its equivalent time-dependent approach.

\section{Discussions}

In this section we discuss a relation between STDDM and a more elaborate extended RPA theory (ERPA)
which has hermiticity \cite{TS}. First we briefly present the ERPA theory. 
The ground state $|\Phi_0\rangle$ in ERPA is constructed so that
\begin{eqnarray}
\langle\Phi_0|[H,a^+_{\alpha}a_{\alpha'}]|\Phi_0\rangle =0,
\label{grc1}
\end{eqnarray}
\begin{eqnarray}
\langle\Phi_0|[H,a^+_{\alpha_1}a^+_{\alpha_2}a_{\alpha_2'}a_{\alpha_1'}]|\Phi_0\rangle =0,
\label{grc2}
\end{eqnarray}
\begin{eqnarray}
\langle\Phi_0|[H,a^+_{\alpha_1}a^+_{\alpha_2}a^+_{\alpha_3}a_{\alpha_3'}a_{\alpha_2'}a_{\alpha_1'}]|\Phi_0\rangle =0
\label{grc3}
\end{eqnarray}
are satisfied for any single-particle indices. 
In other words 
$n^0_{\alpha\alpha'}$, 
$C^0_{\alpha_1\alpha_2\alpha_1'\alpha_2'}$ and the three-body correlation matrix
$C^0_{\alpha_1\alpha_2\alpha_3\alpha_1'\alpha_2'\alpha_3'}$
are determined so that the above  equations are satisfied.
The explicit expressions for Eqs.(\ref{grc1})-(\ref{grc3}) are shown in Ref.\cite{TS}.
Although it is not evident to find an
analytic solution of Eqs.(\ref{grc1})-(\ref{grc3}) \cite{TSW},
a method for obtaining $n^0_{\alpha\alpha'}$
and $C^0_{\alpha_1\alpha_2\alpha_1'\alpha_2'}$ numerically
has been proposed in Ref.\cite{T95} and already been tested for realistic nuclei \cite{T01,T02,T98}
as was shown in Sect.3.
The ERPA equations are formulated using the equation of motion
approach \cite{Rowe,Ring} as 
\begin{eqnarray}
\langle\Psi_0|[[a^+_{\alpha}a_{\alpha'},H],Q^+]|\Psi_0\rangle
=
\omega\langle\Psi_0|[a^+_{\alpha}a_{\alpha'},Q^+]|\Psi_0\rangle
\label{erpa1}
\end{eqnarray}
\begin{eqnarray}
\langle\Psi_0|[[:a^+_{\alpha_1}a^+_{\alpha_2}a_{\alpha_2'}a_{\alpha_1'}:,H],Q^+]|\Psi_0\rangle
=
\omega\langle\Psi_0|[:a^+_{\alpha_1}a^+_{\alpha_2}a_{\alpha_2'}a_{\alpha_1'}:,Q^+]|\Psi_0\rangle,
\label{erpa2}
\end{eqnarray}
where the operator $Q^+$ is defined as
\begin{eqnarray}
\hspace{0.5cm}Q^+=\sum(x_{\lambda\lambda'}a^+_{\lambda}a_{\lambda'}
+X_{\lambda_1\lambda_2\lambda_1'\lambda_2'}:a^+_{\lambda_1}a^+_{\lambda_2}a_{\lambda_2'}a_{\lambda_1'}:),
\label{q-opr}
\end{eqnarray}
and $|\Psi_0\rangle$ is assumed to have the following properties
\begin{eqnarray}
Q^+|\Psi_0\rangle&=&|\Psi\rangle\\
Q|\Psi_0\rangle&=&0.
\end{eqnarray} 
In Eqs.(\ref{erpa2}) and (\ref{q-opr}), $:~:$ stands for 
$:a^+_{\alpha_1}a^+_{\alpha_2}a_{\alpha_2'}a_{\alpha_1'}:=
a^+_{\alpha_1}a^+_{\alpha_2}a_{\alpha_2'}a_{\alpha_1'}-{\cal A}(a^+_{\alpha_1}a_{\alpha_1'}
\langle\Phi_0|a^+_{\alpha_2}a_{\alpha_2'}|\Phi_0\rangle + a^+_{\alpha_2}a_{\alpha_2'}
\langle\Phi_0|a^+_{\alpha_1}a_{\alpha_1'}|\Phi_0\rangle)$.
Equations (\ref{erpa1}) and (\ref{erpa2}) can be written in matrix form
\begin{eqnarray}
\left(
\begin{array}{cc}
A & C \\
B & D
\end{array}
\right)
\left(
\begin{array}{c}
x \\
X
\end{array}
\right)
=\omega
\left(
\begin{array}{cc}
S_1 & T_1 \\
T_2 & S_2
\end{array}
\right)
\left(
\begin{array}{c}
x \\
X
\end{array}
\right) ,
\label{erpa3}
\end{eqnarray}
where each matrix element is evaluated using the ground state $|\Phi_0\rangle$
\begin{eqnarray}
S_1(\alpha'\alpha:\lambda\lambda')
&=&\langle\Phi_0|[a^+_{\alpha}a_{\alpha'},a^+_{\lambda}a_{\lambda'}]|\Phi_0\rangle,
\label{S1}
\\
S_2(\alpha_1'\alpha_2'\alpha_1\alpha_2:\lambda_1\lambda_2\lambda_1'\lambda_2')
&=&\langle\Phi_0|[
:a^+_{\alpha_1}a^+_{\alpha_2}a_{\alpha_2'}a_{\alpha_1'}:,:a^+_{\lambda_1}a^+_{\lambda_2}a_{\lambda_2'}a_{\lambda_1'}:]|\Phi_0\rangle,
\\
T_1(\alpha'\alpha:\lambda_1\lambda_2\lambda_1'\lambda_2')
&=&\langle\Phi_0|[
a^+_{\alpha}a_{\alpha'},:a^+_{\lambda_1}a^+_{\lambda_2}a_{\lambda_2'}a_{\lambda_1'}:]|\Phi_0\rangle ,
\\
T_2(\alpha_1'\alpha_2'\alpha_1\alpha_2:\lambda\lambda')
&=&\langle\Phi_0|[
:a^+_{\alpha_1}a^+_{\alpha_2}a_{\alpha_2'}a_{\alpha_1'}:,a^+_{\lambda}a_{\lambda'}]|\Phi_0\rangle,
\label{T2}
\\
A(\alpha'\alpha:\lambda\lambda')&=&\langle\Phi_0|[
[a^+_{\alpha}a_{\alpha'},H],a^+_{\lambda}a_{\lambda'}]|\Phi_0\rangle,
\\
B(\alpha_1'\alpha_2'\alpha_1\alpha_2:\lambda\lambda')
&=&\langle\Phi_0|[
[:a^+_{\alpha_1}a^+_{\alpha_2}a_{\alpha_2'}a_{\alpha_1'}:,H],a^+_{\lambda}a_{\lambda'}]|\Phi_0\rangle,
\\
C(\alpha'\alpha:\lambda_1\lambda_2\lambda_1'\lambda_2')
&=&\langle\Phi_0|[
[a^+_{\alpha}a_{\alpha'},H],:a^+_{\lambda_1}a^+_{\lambda_2}a_{\lambda_2'}a_{\lambda_1'}:]|\Phi_0\rangle,
\\
D(\alpha_1'\alpha_2'\alpha_1\alpha_2:\lambda_1\lambda_2\lambda_1'\lambda_2')
&=&\langle\Phi_0|[
[:a^+_{\alpha_1}a^+_{\alpha_2}a_{\alpha_2'}a_{\alpha_1'}:,H],
:a^+_{\lambda_1}a^+_{\lambda_2}a_{\lambda_2'}a_{\lambda_1'}:]|\Phi_0\rangle .
\end{eqnarray}
All matrices in the above are written in terms of
$n^0_{\alpha\alpha'}$, $C^0_{\alpha_1\alpha_2\alpha_1'\alpha_2'}$ and 
$C^0_{\alpha_1\alpha_2\alpha_3\alpha_1'\alpha_2'\alpha_3'}$, which are shown in Ref.\cite{TS}. 
Due to Eqs.(\ref{grc1})-(\ref{grc3}), the above matrices have the following symmetries
\begin{eqnarray}
A(\alpha'\alpha:\lambda\lambda')&=&A(\lambda'\lambda:\alpha\alpha')=A(\lambda\lambda':\alpha'\alpha)^*,
\\
B(\alpha_1'\alpha_2'\alpha_1\alpha_2:\lambda\lambda')&=&C(\lambda'\lambda:\alpha_1\alpha_2\alpha_1'\alpha_2')
=C(\lambda\lambda':\alpha_1'\alpha_2'\alpha_1\alpha_2)^*,
\\
D(\alpha_1'\alpha_2'\alpha_1\alpha_2:\lambda_1\lambda_2\lambda_1'\lambda_2')
&=&D(\lambda_1'\lambda_2'\lambda_1\lambda_2:\alpha_1\alpha_2\alpha_1'\alpha_2')
=D(\lambda_1\lambda_2\lambda_1'\lambda_2':\alpha_1'\alpha_2'\alpha_1\alpha_2)^*.
\end{eqnarray}
We explain in more detail the last relation as an example. The following relation always 
holds due to the operator identity
\begin{eqnarray}
D(\alpha_1'\alpha_2'\alpha_1\alpha_2:\lambda_1\lambda_2\lambda_1'\lambda_2')
-D(\lambda_1'\lambda_2'\lambda_1\lambda_2:\alpha_1\alpha_2\alpha_1'\alpha_2')
&=&\langle\Phi_0|[
[:a^+_{\alpha_1}a^+_{\alpha_2}a_{\alpha_2'}a_{\alpha_1'}:,H],
:a^+_{\lambda_1}a^+_{\lambda_2}a_{\lambda_2'}a_{\lambda_1'}:]|\Phi_0\rangle 
\nonumber \\
&-&\langle\Phi_0|[
[:a^+_{\lambda_1}a^+_{\lambda_2}a_{\lambda_2'}a_{\lambda_1'}:,H],
:a^+_{\alpha_1}a^+_{\alpha_2}a_{\alpha_2'}a_{\alpha_1'}:]|\Phi_0\rangle
\nonumber \\
&=&\langle\Phi_0|
[H,[:a^+_{\alpha_1}a^+_{\alpha_2}a_{\alpha_2'}a_{\alpha_1'}:,
:a^+_{\lambda_1}a^+_{\lambda_2}a_{\lambda_2'}a_{\lambda_1'}:]]|\Phi_0\rangle. 
\label{D}
\end{eqnarray}
The commutation relation between two-body operators in the last line of the above equation
is reduced to three-body operators at most.  Since Eqs.(\ref{grc1})-(\ref{grc3}) are valid,
the last line of Eq.(\ref{D}) is identical to zero. Thus, the symmetry of the matrix $D$ is proven.

The ortho-normality condition in ERPA is given by \cite{Taka}
\begin{eqnarray}
(x^*_{\mu'}~ X^*_{\mu'})\left(
\begin{array}{cc}
S_1 & T_1 \\
T_2& S_2
\end{array}
\right)
\left(
\begin{array}{c}
x_{\mu} \\
X_{\mu}
\end{array}
\right)
=\delta_{\mu\mu'},
\end{eqnarray}
where $x_{\mu}$ and $X_{\mu}$ constitute eigenstates of Eq.(\ref{erpa3}) with $\omega=\omega_{\mu}$.
The completeness relation becomes
\begin{eqnarray}
\sum_{\mu}
\left(
\begin{array}{c}
x_{\mu} \\
X_{\mu}
\end{array}
\right)
(x^*_{\mu} X^*_{\mu})
\left(
\begin{array}{cc}
S_1 & T_1 \\
T_2& S_2
\end{array}
\right)
=I .
\end{eqnarray} 
The transition amplitudes for one-body and two-body operators, 
$z_{\alpha\alpha'}=\langle\Psi_0|a^+_{\alpha'}a_{\alpha}|\Psi\rangle$
and $Z_{\alpha_1\alpha_2\alpha_1'\alpha_2'}=\langle\Psi_0|:a^+_{\alpha_1'}a^+_{\alpha_2'}a_{\alpha_1}a_{\alpha_2}:|\Psi\rangle$,
are calculated as follows
\begin{eqnarray}
\left(
\begin{array}{c}
z \\
Z
\end{array}
\right)=
\left(
\begin{array}{cc}
S_1 & T_1 \\
T_2& S_2
\end{array}
\right)
\left(
\begin{array}{c}
x \\
X
\end{array}
\right).
\end{eqnarray} 

Now we show a relation between STDDM and ERPA. When the eigenvector $(x, X)$ in STDDM is transformed to $(y, Y)$ as
\begin{eqnarray}
\left(
\begin{array}{c}
x \\
X
\end{array}
\right)=
\left(
\begin{array}{cc}
S_1 & T_1 \\
T_2& S'_2
\end{array}
\right)
\left(
\begin{array}{c}
y \\
Y
\end{array}
\right),
\end{eqnarray}
where $S'_2$ has no terms with the three-body correlation matrices,
the equation in STDDM becomes
\begin{eqnarray}
\left(
\begin{array}{cc}
aS_1+cT_2 & aT_1+cS'_2 \\
bS_1+dT_2 & bT_1+dS'_2
\end{array}
\right)
\left(
\begin{array}{c}
y \\
Y
\end{array}
\right)
=\omega
\left(
\begin{array}{cc}
S_1 & T_1 \\
T_2 & S'_2
\end{array}
\right)
\left(
\begin{array}{c}
y \\
Y
\end{array}
\right).
\label{erpa4}
\end{eqnarray}
It is straightforward to prove that when the three-body correlation matrices
$C_{\alpha_1\alpha_2\alpha_3\alpha_1'\alpha_2'\alpha_3'}$ are neglected,
three blocks of the above hamiltonian matrix satisfy the identities
$A=aS_1+cT_2$, $B=bS_1+dT_2$, and $C=aT_1+cS'_2$.
We explain the relation $B=bS_1+dT_2$ in detail.
When two-body and three-body operators are decomposed using 
$a^+_{\alpha_1}a^+_{\alpha_2}a_{\alpha_2'}a_{\alpha_1'}=:a^+_{\alpha_1}a^+_{\alpha_2}a_{\alpha_2'}a_{\alpha_1'}:
+{\cal A}(a^+_{\alpha_1}a_{\alpha_1'}
\langle\Phi_0|a^+_{\alpha_2}a_{\alpha_2'}|\Phi_0\rangle+a^+_{\alpha_2}a_{\alpha_2'}
\langle\Phi_0|a^+_{\alpha_1}a_{\alpha_1'}|\Phi_0\rangle)$ 
for a two-body operator and a similar relation for a three-body operator,
the commutation  relation $[:a^+_{\alpha_1}a^+_{\alpha_2}a_{\alpha_2'}a_{\alpha_1'}:,H]$
becomes
\begin{eqnarray}
[:a^+_{\alpha_1}a^+_{\alpha_2}a_{\alpha_2'}a_{\alpha_1'}:,H]&=&
\sum_{\lambda\lambda'}b(\alpha_1\alpha_2\alpha_1'\alpha_2':\lambda\lambda')a^+_{\lambda'}a_{\lambda}
\nonumber \\
&+&\sum_{\lambda_1\lambda_2\lambda_1'\lambda_2'}d(\alpha_1\alpha_2\alpha_1'\alpha_2':\lambda_1\lambda_2\lambda_1'\lambda_2')
:a^+_{\lambda_1'}a^+_{\lambda_2'}a_{\lambda_2}a_{\lambda_1}:
\nonumber \\
&+&\sum_{\lambda_1\lambda_2\lambda_3\lambda_1'\lambda_2'\lambda_3'}
e(\alpha_1\alpha_2\alpha_1'\alpha_2':\lambda_1\lambda_2\lambda_3\lambda_1'\lambda_2'\lambda_3')
:a^+_{\lambda_1'}a^+_{\lambda_2'}a^+_{\lambda_3'}a_{\lambda_3}a_{\lambda_2}a_{\lambda_1}:,
\label{comm}
\end{eqnarray}
where the matrix $e$ consists of matrix elements of the two-body interaction. In STDDM, the last term on the right-hand side
of the above equation is neglected. It is clear from Eq.(\ref{comm}) that 
$B=\langle\Phi_0|[
[:a^+_{\alpha_1}a^+_{\alpha_2}a_{\alpha_2'}a_{\alpha_1'}:,H],a^+_{\lambda}a_{\lambda'}]|\Phi_0\rangle
=bS_1+dT_2$.
The matrix $D$ has some terms which cannot be written by using $bT_1+dS'_2$
even if the three-body correlation matrices are neglected:
For example, all terms with the square of the two-body correlation matrix
cannot be expressed with $bT_1$.
Thus it is found that STDDM is identical to ERPA under the following two assumptions: The first is that
the three-body correlation matrices can be neglected and the second is that $D\approx bT_1+dS'_2$.
Our numerical calculations so far performed suggest that these assumptions are quite reasonable.

\section{Summary}
The eigenstates of STDDM were calculated for the low-lying $2^+$ states in
$^{24}$O to better understand the foundation of its corresponding time-dependent approach.
It is found that STDDM properly deals with the effects of ground-state correlations
on the low-lying $2^+$ states and that the non-hermiticity of STDDM is quite moderate:
The eigenvalues have quite small imaginary parts and the strength function is practically positive
definite although it is not guaranteed because of its non-hermitian form.
A comparison of STDDM and an extended RPA (ERPA) theory with hermiticity was also made, and it is found that STDDM 
is a reasonable approximation to the ERPA theory.

\appendix
\section{}
The matrices in Eq.(\ref{stddm1}) are given by:
\begin{eqnarray}
a(\alpha\alpha':\lambda\lambda')&=&(\epsilon_{\alpha}-\epsilon_{\alpha'})\delta_{\alpha\lambda}\delta_{\alpha'\lambda'}
-\sum_{\beta}(\langle\beta\lambda'|v|\alpha'\lambda\rangle_An^0_{\alpha\beta}
-\langle\alpha\lambda'|v|\beta\lambda\rangle_An^0_{\beta\alpha'}),
\\
b(\alpha_1\alpha_2\alpha_1'\alpha_2':\lambda\lambda')&=&
-\delta_{\alpha_1\lambda}\{\sum_{\beta\gamma\delta}[(\delta_{\alpha_2\beta}-n^0_{\alpha_2\beta})
n^0_{\gamma\alpha_1'}n^0_{\delta\alpha_2'}
+n^0_{\alpha_2\beta}(\delta_{\gamma\alpha_1'}-n^0_{\gamma\alpha_1'})(\delta_{\delta\alpha_2'}-n^0_{\delta\alpha_2'})]
\langle\lambda'\beta|v|\gamma\delta\rangle_A
\nonumber \\
&+&\sum_{\beta\gamma}(\langle\lambda'\alpha_2|v|\beta\gamma\rangle C^0_{\beta\gamma\alpha_1'\alpha_2'}
+\langle\lambda'\beta|v|\alpha_1'\gamma\rangle_A C^0_{\alpha_2\gamma\alpha_2'\beta}
-\langle\lambda'\beta|v|\alpha_2'\gamma\rangle_A C^0_{\alpha_2\gamma\alpha_1'\beta}\}
\nonumber \\
&+&\delta_{\alpha_2\lambda}\{\sum_{\beta\gamma\delta}[(\delta_{\alpha_1\beta}-n^0_{\alpha_1\beta})
n^0_{\gamma\alpha_1'}n^0_{\delta\alpha_2'}
+n^0_{\alpha_1\beta}(\delta_{\gamma\alpha_1'}-n^0_{\gamma\alpha_1'})(\delta_{\delta\alpha_2'}-n^0_{\delta\alpha_2'})]
\langle\lambda'\beta|v|\gamma\delta\rangle_A
\nonumber \\
&+&\sum_{\beta\gamma}[\langle\lambda'\alpha_1|v|\beta\gamma\rangle C^0_{\beta\gamma\alpha_1'\alpha_2'}
+\langle\lambda'\beta|v|\alpha_1'\gamma\rangle_A C^0_{\alpha_1\gamma\alpha_2'\beta}
-\langle\lambda'\beta|v|\alpha_2'\gamma\rangle_A C^0_{\alpha_1\gamma\alpha_1'\beta}]\}
\nonumber \\
&+&\delta_{\alpha_1'\lambda'}\{\sum_{\beta\gamma\delta}[(\delta_{\delta\alpha_2'}-n^0_{\delta\alpha_2'})
n^0_{\alpha_1\beta}n^0_{\alpha_2\gamma}
+n^0_{\delta\alpha_2'}
(\delta_{\alpha_1\beta}-n^0_{\alpha_1\beta})(\delta_{\alpha_2\gamma}-n^0_{\alpha_2\gamma})]
\langle\beta\gamma|v|\lambda\delta\rangle_A
\nonumber \\
&+&\sum_{\beta\gamma}[\langle\beta\gamma|v|\lambda\alpha_2'\rangle C^0_{\alpha_1\alpha_2\beta\gamma}
+\langle\alpha_1\beta|v|\lambda\gamma\rangle_A C^0_{\alpha_2\gamma\alpha_2'\beta}
-\langle\alpha_2\beta|v|\lambda\gamma\rangle_A C^0_{\alpha_1\gamma\alpha_2'\beta}]\}
\nonumber \\
&-&\delta_{\alpha_2'\lambda'}\{\sum_{\beta\gamma\delta}[(\delta_{\delta\alpha_1'}-n^0_{\delta\alpha_1'})
n^0_{\alpha_1\beta}n^0_{\alpha_2\gamma}
+n^0_{\delta\alpha_1'}
(\delta_{\alpha_1\beta}-n^0_{\alpha_1\beta})(\delta_{\alpha_2\gamma}-n^0_{\alpha_2\gamma})]
\langle\beta\gamma|v|\lambda\delta\rangle_A
\nonumber \\
&+&\sum_{\beta\gamma}[\langle\beta\gamma|v|\lambda\alpha_1'\rangle C^0_{\alpha_1\alpha_2\beta\gamma}
+\langle\alpha_1\beta|v|\lambda\gamma\rangle_A C^0_{\alpha_2\gamma\alpha_1'\beta}
-\langle\alpha_2\beta|v|\lambda\gamma\rangle_A C^0_{\alpha_1\gamma\alpha_1'\beta}]\}
\nonumber \\
&+&\sum_{\beta}[\langle\alpha_1\lambda'|v|\beta\lambda\rangle_A C^0_{\beta\alpha_2\alpha_1'\alpha_2'}
-\langle\alpha_2\lambda'|v|\beta\lambda\rangle_A C^0_{\beta\alpha_1\alpha_1'\alpha_2'}
\nonumber \\
&-&\langle\beta\lambda'|v|\alpha_2'\lambda\rangle_A C^0_{\alpha_1\alpha_2\alpha_1'\beta}
+\langle\beta\lambda'|v|\alpha_1'\lambda\rangle_A C^0_{\alpha_1\alpha_2\alpha_2'\beta}],
\\
c(\alpha\alpha':\lambda_1\lambda_2\lambda_1'\lambda_2')&=&
\langle\alpha\lambda_2'|v|\lambda_1\lambda_2\rangle\delta_{\alpha'\lambda_1'}
-\langle\lambda_1'\lambda_2'|v|\alpha'\lambda_2\rangle\delta_{\alpha\lambda_1},
\\
d(\alpha_1\alpha_2\alpha_1'\alpha_2':\lambda_1\lambda_2\lambda_1'\lambda_2')&=&
(\epsilon_{\alpha_1}+\epsilon_{\alpha_2}-\epsilon_{\alpha_1'}-\epsilon_{\alpha_2'})
\delta_{\alpha_1\lambda_1}\delta_{\alpha_2\lambda_2}
\delta_{\alpha_1'\lambda_1'}\delta_{\alpha_2'\lambda_2'}
\nonumber \\
&+&\delta_{\alpha_1'\lambda_1'}\delta_{\alpha_2'\lambda_2'}
\sum_{\beta\gamma}(\delta_{\alpha_1\beta}\delta_{\alpha_2\gamma}
-\delta_{\alpha_2\gamma}n^0_{\alpha_1\beta}
-\delta_{\alpha_1\beta}n^0_{\alpha_2\gamma})\langle\beta\gamma|v|\lambda_1\lambda_2\rangle
\nonumber \\
&-&\delta_{\alpha_1\lambda_1}\delta_{\alpha_2\lambda_2}
\sum_{\beta\gamma}(\delta_{\alpha_1'\beta}\delta_{\alpha_2'\gamma}
-\delta_{\alpha_2'\gamma}n^0_{\beta\alpha_1'}
-\delta_{\alpha_1'\beta}n^0_{\gamma\alpha_2'})
\langle\lambda_1'\lambda_2'|v|\beta\gamma\rangle
\nonumber \\
&+&\delta_{\alpha_2\lambda_2}\delta_{\alpha_2'\lambda_2'}
\sum_{\beta}(\langle\alpha_1\lambda_1'|v|\beta\lambda_1\rangle_An^0_{\beta\alpha_1'}
-\langle\beta\lambda_1'|v|\alpha_1'\lambda_1\rangle_An^0_{\alpha_1\beta})
\nonumber \\
&+&\delta_{\alpha_2\lambda_2}\delta_{\alpha_1'\lambda_1'}
\sum_{\beta}(\langle\alpha_1\lambda_2'|v|\beta\lambda_1\rangle_An^0_{\beta\alpha_2'}
-\langle\beta\lambda_2'|v|\alpha_2'\lambda_1\rangle_An^0_{\alpha_1\beta})
\nonumber \\
&+&\delta_{\alpha_1\lambda_1}\delta_{\alpha_1'\lambda_1'}
\sum_{\beta}(\langle\alpha_2\lambda_2'|v|\beta\lambda_2\rangle_An^0_{\beta\alpha_2'}
-\langle\beta\lambda_2'|v|\alpha_2'\lambda_2\rangle_An^0_{\alpha_2\beta})
\nonumber \\
&+&\delta_{\alpha_1\lambda_1}\delta_{\alpha_2'\lambda_2'}
\sum_{\beta}(\langle\alpha_2\lambda_1'|v|\beta\lambda_2\rangle_An^0_{\beta\alpha_1'}
-\langle\beta\lambda_1'|v|\alpha_1'\lambda_2\rangle_An^0_{\alpha_2\beta}),
\end{eqnarray}
where $n^0_{\alpha\alpha'}=\langle\Phi_0|a^+_{\alpha'}a_\alpha|\Phi_0\rangle$ and
$C^0_{\alpha_1\alpha_2\alpha_1'\alpha_2'}=
\langle\Phi_0|a^+_{\alpha_1'}a^+_{\alpha_2'}a_{\alpha_2}a_{\alpha_1}|\Phi_0\rangle
-{\cal A}(n^0_{\alpha_1\alpha_1'}n^0_{\alpha_2\alpha_2'})$, and
the subscript $A$ indicates that the corresponding matrix element is antisymmetrized.

\end{document}